\documentclass{article}
\usepackage{amssymb}
\usepackage{amsmath}

\newtheorem{Th}{Corollary}

\begin{document}

\title{COMMENT ON A TONOMURA EXPERIMENT : LOCALITY OF THE VECTOR POTENTIAL}         
\author{OLIVIER COSTA DE BEAUREGARD AND GEORGES LOCHAK\footnote{Fondation LOUIS DE BROGLIE 23 rue MARSOULAN 75012 PARIS}}        
\date{}          
\maketitle
\begin{abstract}
Three predictions for additional tests in a Tonomura experiment: 1,2: The Fresnel frin-ges displayed outside and inside the geometric shadow of a toroidal magnet should subsist intact, the ones if the others are masked, and vice versa ; 3 : Placing the registering film just before the magnet and thus uncovering the entire fringe pattern  should display the curved fringes connecting the outer and inner straight ones. Physicality of the vector potential expressed in the source adhering gauge will thus be unequivocally proved.   
\end{abstract}

\section{An unconventional testable claim}   
De Broglie has tersely stated~\cite{broglie1} that his~\cite{broglie2} universal formula 
\begin{equation}
P^{i}\equiv \mu U^{i} -eA^{i} = \hbar k^{i}                                             
\end{equation}

relating the canonical $4$-momentum $P^{i}$ of a point charge of charge $-e$ , rest mass $\mu$, $4$-velocity $U^{i}$, to the $4$-frequency $k{i}$ of the associated wave, selects uniquely the electromagnetic gauge. The point is : In absence of external electromagnetic sources, adding to the $4$-potential $A^{i}$ an arbitrary $4$-gradient would entail indefiniteness of the $4$-frequency -a cardiac arythmy of the electron, so to speak. This is not observed -and is denied by crystal diffraction unequivocally displaying (in standard notation) the formula
\begin{equation}
\bf{p} \equiv mv = \hbar \bf{k}                                                    
\end{equation}
  What then of gauge invariance of the Dirac equation? Adding to the canonical $4$-momentum operator $i \hbar\partial_{i} -eA_{i}$ an arbitrary $4$-gradient can be compensated by substracting this same $4$-gradient from the wave function's phase. All right -this is like cashing a cheque.
  An invariance law of a differerential equation need not subsist in its solutions, which imply integration conditions. What de Broglie means is that in the expression of a free electron's $4$-momentum the $4$-potential is identically zero : $Ai \equiv 0$ in absence of electromagnetic sources. This is unquestionable.     
 \begin{Th}
  Any sort of electron interference experiment performed in presence of a toroidal magnet displays the curlless vector potential ${\bf{A}}(r)$ as expressed in the source adhering gauge -this being tantamount to a measurement of the vector potential. 
\end{Th}
  So we claim (a big step forward !) that : A locally observable effect underlies the A.B. effect. 

\section{Proof via a Tonomura experiment}   
Tonomura~\cite{tonomura} has combined an 'electron biprism inter-ference' with an Aharonov-Bohm one. A very perfect toroidal magnet of trapped flux $\Phi$ quantized in $h/2e$ units~\cite{costa1} placed downstream of a 'biprism' has its axis z orthogonal to the planes displaying 'normally' the Fresnel fringes. A registering film, placed 'normally' after the magnet, displays outside and inside its circular shadow straight Fresnel fringes which are either identical to each other or black-to-white exchanged, depending on the flux $\Phi$ being an even or an odd multiple of $h/2e$. The fact is that the magnet's shadow, now termed the black ring, is 'geometric' style showing no circular fringes, and thus no explicit $A.B$. effect. This precludes any observable interference between the external and internal fringes -which can be tested. \\ \\
  Obturating the inside of the black ring should not affect in any way the external fringes which, displayed as parallel straight lines, are identical to those existing in absence of the magnet -because the magnet's influence is asymptotically zero. Slipping transversally the magnet out of the picture will just uncover the genuine Fresnel fringes. 
  Similarly, for the reasons stated, obturating the outside of the black ring should not affect in any way the internal  fringes -a very crucial test of locality ! \\
  How the outside and inside fringes are linked together is hidden by the black ring. Between them exists a phase shift amounting to a multiple of $h/e$, due to addition of $-e\bf{A}$ to the two kinetic momenta $mv $combined with obliquity of the two interfering $\bf{k}$ vectors vizz the axis $z$. 
  The hidden curved fringes can be recovered by placing the registering film just before the magnet, thus wiping off the black ring. The fringe pattern displayed will not be the genuine Fresnel one, but one where curved fringes now connect the external and internal straight Tonomura ones -a very strong proof of local physicality of $\bf{A}$ ahead of electron impact ! 
\section{Direct mesurement of the vector potential}
 If the vector potential $\bf{A}$ is a locally measurable magnitude a precise measurement of a curlless vector potential is possible. Rather than a spatially extended 'electron biprism' one should then use as interference generator a small diffracting crystal. \\
  Performed inside the curlless vector potential $\bf{A}(r)$ generated by a toroidal magnet,  crystal diffraction will evidence, instead of formula (2),  the formula
\begin{equation}
\hbar \bf{k} = m \bf{v} - e \bf{A}                                                    
\end{equation}

yielding a measurement of $\bf{A}$ expressed in the source adhering gauge.\\ 
  The maximal and neater effect will obtain if the magnet's center coincides with that of the crystal and its axis with that of the gun. Then turning the magnet around its center will modify the intensity along the circular rings.

\section{Resurrection of the potentials 'assassinated' by Heaviside}
 Electromagnetic gauge invar-iance states : Forces, linear or angular, depend on the fields, not the potentials. All right, this is very true.\\
  But the integrals of forces -over space, energies, or over time, momenta (linear or angular ; the 6-component angular momentum including the boost) do depend on the potentials. As interaction energies and linear or angular momenta of bound systems are measurable mag-nitudes the attached potentials also are -with expressions selected as integration conditions~\cite{costa1}.\\
  This is well known but underestimated in the case of Einstein's energy-mass equivalence : the electrostatic mass defect of a bound system is part of its total mass.\\
  By relativistic covariance there follows~\cite{costa1} that action-reaction (linear or angular) also selects the source adhering gauge as an integration condition ; an example is afforded by the Wheeler-Feynman electrodynamics. \\
  Electromagnetically induced inertia thus emerges as a general concept to be discussed elswhere. \\
  De Broglie~\cite{broglie1}~\cite{costa2} has stated that both the Einstein $W=c^{2}m$ energy-mass and the Planck $W=h\nu$ energy-frequency equivalences select the electromagnetic gauge ; covariant expressions of these statements have been produced~\cite{costa2}. 

\end{document}